# Electro-optical Neural Networks based on Time-stretch Method

Yubin Zang, Minghua Chen, *Member, IEEE*, Sigang Yang, *Member, IEEE* and Hongwei Chen, *Senior Member, IEEE*

*Abstract*—In this paper, a novel architecture of electro-optical neural networks based on the time-stretch method is proposed and numerically simulated. By stretching time-domain ultrashort pulses, multiplications of large scale weight matrices and vectors can be implemented on light and multiple-layer of feedforward neural network operations can be easily implemented with fiber loops. Via simulation, the performance of a three-layer electro-optical neural network is tested by the handwriting digit recognition task and the accuracy reaches 88% under considerable noise.

*Index Terms*—Electro-optical neural networks, time-stretch method, handwriting digit recognition

## I. INTRODUCTION

Artificial intelligence (AI) has been in rapid and profound development in recent years. Neural network, as one of the most frequently and widely used hierarchies in the field of AI, has undergone great progress thanks to the development of integrated circuits. So far, electronic neural networks have had great applications in fields of audio signal processing, digital image processing and pattern recognition etc. By adapting and optimizing learning algorithms, electronic neural networks can reach high accuracy in both regression and prediction tasks with a fast convergence rate. For instance, electronic neural networks can reach over 95% accuracy in handwriting digit recognition from Modified National Institute of Standards and Technology (MNIST) test set [1], [2]. CPU, GPU, FPGA and ASIC are the four main platforms for operating electronic neural networks [1].

Even though electronic neural networks (ENN) have developed relatively fast in recent years, they still have several drawbacks which are hard to overcome. First of all, ENN will cause relatively large time delay in most cases when operating. This is because their large weight coefficients are stored in memory modules of electronic devices [3]. Therefore, it will take a lot of time to transmit weight coefficients from memory modules to processing units before computation and transmit results from processing units to memory modules after computation. Secondly, power efficiency will be relatively low since most power is used to exchange data between processing units and memory modules. Moreover, as the scale of neural networks expands, especially for implementing complicated deep learning neural networks, electronic devices with large bandwidth are in urgently need to transmit signals as fast as possible. However, these devices are hard to manufacture.

Instead of using electron as the carrier for signal transmitting and processing, optical neural networks which use photon as the information carrier have been put forward. Compared with traditional electronic neural networks, optical neural networks can fundamentally overcome the known challenges. Firstly, there will be less time delay in optical neural networks since their weight coefficients are directly related with parameters and properties of light or photonic devices. After computation, results will be immediately processed and brought to the next layer of neural networks at the speed of light. Secondly, power efficiency can be relatively higher than the electronic counterparts since there is no need to use power to transmit information between memory modules and computing units. Last but not least, thanks to high-speed optical fiber communication technique, information between different layers of neural networks or even between different optical neural networks will be transmitted at quite a high speed.

Over the past 20 years, many architectures of optical neural networks have been proposed. The most important architecture for chip-integrated optical neural networks was first proposed by M. Reck and *et al* in 1994 [4]. In this article, they proposed a triangular-shape structure of multiport interferometers whose transfer matrix is unitary matrix. This structure which is usually implemented by MZI networks, together with tunable attenuator networks, can later be used as linear computing units in optical neural networks. However, it will be relatively unstable when the scale of the structure expands or when taking into the account of the attenuation effect [4], [5]. Based on Reck's structure, W. R. Clements and *et al.* from the University of Oxford proposed another optimized rectangular-shape structure in realizing the same unitary matrix [5]. This new structure and its corresponding decomposition method overcomes the problems in Reck's design and thus is more robust [5]. In 2017, Y. Shen and *et al.* from MIT successfully proposed a chip-integrated phonetic neural network by adapting the structure proposed by Reck as its linear computing

The work of this manuscript was supported by Beijing Municipal Science & Technology Commission with Grant No. Z181100008918011, NSFC under Contract 61771284 and Beijing Natural Science Foundation under Contracts L182043.

Yubin Zang, Minghua Chen, Sigang Yang, Hongwei Chen are with Beijing National Research Center for Information Science and Technology (BNRist), Department of Electronic Engineering, Tsinghua University, Beijing, 100084, China (E-mail: chenhw@tsinghua.edu.cn).

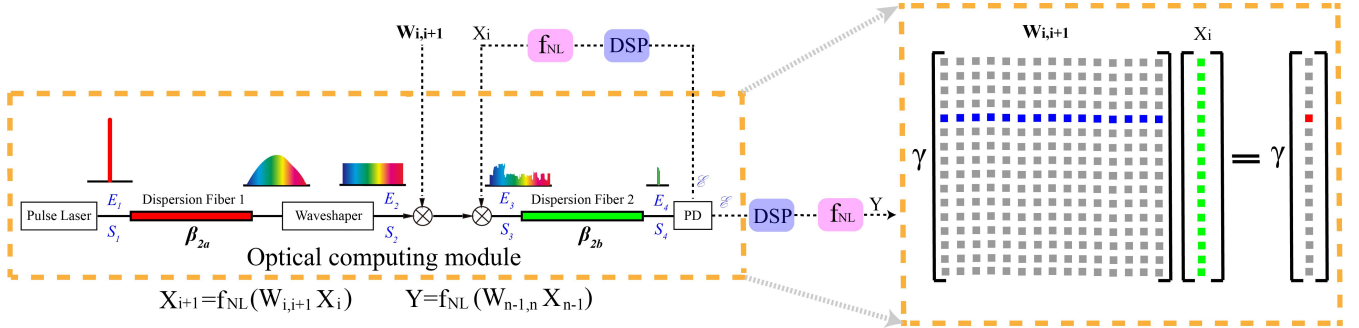

Fig. 1. Universal structure and principle of TS-NN. Main picture: loop-based structure of TS-NN which mainly consists of pulse laser, dispersion fiber 1, waveshaper, two modulators, dispersion fiber 2, PD, DSP and nonlinear module. Solid black arrows and lines in the picture show the optical information flow while dashed black arrows and lines show electronic flow. The structure in the window surrounded by yellow dashed line is the core optical structure of TS-NN. For each pulse in each loop, it can accomplish the multiplication of the input vector $X_i$ and one row of the weight matrix $W_{i,i+1}$. This process is described in the magnified window where $\gamma$ is the proportional constant. Note that part of the last output loop-PD, DSP and nonlinear module is unwrapped in this figure.

component. By using the chip four times and simulating the transfer function of saturable absorbers as nonlinear functions, this chip-integrated three-layer optical feedforward neural network can reach 76.7% accuracy in classifying 4 vowels [6]. Diffractive Deep-learning Neural Network ($D^2NN$) structure proposed by X. Lin and *et al* from UCLA in 2018 has cast another view on optical neural networks [7]. Researchers mapped weight coefficients into different thickness at different position of diffractive layers and used diffraction effect of light as a way to implement linear computation. They constructed a feedforward neural network with 5 diffractive layers which achieves 91.75% and 81.13% in handwriting digit recognition task and fashion set classification task respectively [7].

Although fully utilizing photonic instruments to implement both linear and nonlinear part of feedforward neural networks can overcome some of the main drawbacks of conventional ENN, it will cause other problems. Firstly, nonlinear functions are hard to implement by photonic devices. As can be seen from the nanophotonic neural network chip proposed by Y. Shen and *et al* [6], its nonlinear functions which should have been implemented by directly using saturable absorbers, were implemented by numerically simulating their transfer functions instead. For $D^2NN$ structure proposed by X. Lin and *et al* [7], this problem is avoided by crudely eliminating the nonlinear part of neural networks. Secondly, since weight coefficients are pre-set into the properties of photonic instruments, the ability of reconfiguration is relatively low for conventional ONN. For nanophotonic neural network circuits proposed by Y. Shen and *et al* [6], each reconfiguration of weight coefficients will cost relatively long time since this process is implemented by tuning phase shifter based on thermo-optical effect. For $D^2NN$ structure proposed by X. Lin and *et al* [7], its diffractive layers are specifically designed and fabricated for one task. For other task, diffractive layers must be re-designed and re-fabricated which costs extra time and expenses. Besides, conventional optical neural networks are hard to adjust and align and sometimes even fragile for implementing large-scale theoretical model of neural networks, especially for those system in which information are processed and transmitted in a parallel way. For example, for the nanophotonic circuits proposed by Y. Shen and *et al* [6], even for the task of vowel recognition with only 4 inputs and 4 outputs, it is hard to adjust and align the whole system with 56 MZIs. For implementing more complicated tasks with more inputs and outputs such as MNIST digit recognition, face recognition and etc, the scale of MZIs will expend severly which will even make the whole system unreachable.

In order to take both advantages of electronic and photonic systems, electro-optical neural networks have been put forward in recent years. Most of the state of the art electro-optical neural networks are proposed to realize recurrent neural networks (RNN) which are good at time series prediction, language processing and etc. [8]-[10]. Few electro-optical neural networks focus on implementing feedforward neural networks which not only play an important role in image processing, pattern recognition and etc., but also are the basis for convolutional neural networks (CNN).

In this paper, we propose a new structure of electro-optical neural networks based on time-stretch method to implement feedforward neural networks. In our system, linear matrix computations are done with quite a high speed by photonic devices due to their advantages over calculation speed, transmission loss and spectrum bandwidth, while nonlinear functions can be more precisely implemented by electronic devices. Time-stretch, which is the most crucial technology in our electro-optical system, is also an important method for optical signal processing [11] and the method has been widely used in the fields of ultra-fast real-time ADC [12]-[20], analyzing optical noise and fluctuations [21]-[23], ultrashort nonlinear optical phenomena [24]-[31] and imaging [32]-[35]. By adapting the time-stretch method, a relatively large number of coefficients can be modulated subsequently onto the broadened pulses to transmit information from one layer of neural network to the next in a serial way. In this case, not only large scale multi-layer feedforward neural networks can be implemented, but also the total robust and reconfiguration performances of the system can be improved due to the adaption of electronic devices.

The rest of this paper is organized into three parts. In section II, the core optical structure of our time-stretch electro-optical neural network (TS-NN) is firstly introduced and then the crucial statement that by adapting time-stretch method, this core optical structure which mainly includes the pulse generation, broaden, modulation, compression components and photo-diode (PD) can implement linear computations in neural networks is verified theoretically by mathematical analysis. Thus, together with nonlinear stimulus function implemented by electronic devices, multiple machine learning tasks which have been accomplished by ENN before can run on our system. In section III, TS-NN is designed and constructed by simulation and then the performances of this TS-NN are evaluated and discussed. Conclusion and prospects of this TS-NN are discussed in section IV.

## II. Principle and Analysis

### A. Main structure of TS-NN

As is shown in Fig. 1, our system is a loop-based system and the number of loops is related with the number of layers of the feedforward neural network. Specifically, ($N$-$1$) loops are used to implement an $N$-layer neural network. Generally, two kinds of operations—linear computation (matrix multiplication) and nonlinear transform must be done in each loop. The core optical structure which adapts time-stretch method is used to implement the linear computation process. After that, the results will be processed by digital signal processor (DSP) and then go through nonlinear function. At last, the outputs will serve as the modulated input vector in the next loop.

Mode-locked laser is used as the light source to generate ultrashort periodic pulses. By adapting time-stretch method, these pulses can be further broadened by the fiber link with relatively large dispersion and then be reshaped to flatten by the waveshaper. Afterwards, the flattened broadened pulses will be modulated with each row of elements from weight matrix firstly and the input vector which was turned from the outputs of the previous loop secondly. After pulses go through dispersion fiber components and PD devices, energy of each pulse which implies the result of multiplication of each row of elements from the weight matrix and the input vector which was turned from the output results of the previous loop will be accumulated and then be processed by DSP using signal processing algorithm. At last, the results will go through nonlinear stimulus function such as non-negative sigmoid function which is simulated by electronic device in our system to become the input vector of the next loop. After signals go through all loops, the outputs of TS-NN will be further processed and analyzed.

### B. Mathematical analysis of TS-NN

For this section, it is assumed that the pulse generated by mode-locked laser has a rectangular-shape spectrum without any chirp. According to Dispersive Fourier Transform, this assumption is equivalent to the processes that the ultrashort pulse comes from a mode-locked laser and then a waveshaper or programmable optical filter is adapted to turn the broadened pulse into rectangular-shape after propagating through the fiber link as long as the dispersion is large enough to meet the requirement for stationary phase approximation [36].

The spectrum of the pulse can be expressed as

$$S_1(0, \omega - \omega_s) = \frac{2\pi}{\Omega}[\varepsilon(\omega - \omega_s - \frac{\Omega}{2}) - \varepsilon(\omega - \omega_s + \frac{\Omega}{2})] = G_\Omega(\omega - \omega_s) \quad (1)$$

in which $\Omega$ is the spectrum bandwidth and the function $\varepsilon = \varepsilon(x)$ represents the normalized Heaviside function (unit step function). Its value equals to one when $x$ is larger than zero, otherwise it equals to zero. The function $G = G_\Omega(x)$ represents the rectangular-shaped function which is centered in the origin point. Its width and height equal to $\Omega$ and $2\pi/\Omega$ respectively. The function notation $G$ will be further used in formula derivations. $\omega_s$ ensembles the central angular frequency of the light pulse.

According to the theory of systems and signals, the corresponding time domain signal can be calculated precisely by using the definition of the Inverse Fourier Transform [36], [37], [38].

$$E(z,t) = \frac{1}{2\pi} \int_{-\infty}^{\infty} S(z, \omega - \omega_s) \exp\{-j(\omega - \omega_s)t\} d\omega \quad (2)$$

in which $z$ represents the distance of light propagation.

From (1) and (2), the corresponding time-domain signal can be obtained as

$$E_1(0,t) = \frac{\sin(\Omega t/2)}{\Omega t/2} = Sa\left(\frac{\Omega t}{2}\right) \quad (3)$$

where the function $Sa = Sa(x)$ is defined as $Sa(x) = \sin(x)/x$ in the theory of systems and signals [37], [38]. This notation will be used in proceeding derivations as well. This function has one main lobe whose width and height equal to $2\pi$ and 1 respectively with infinite side lobes oscillating along the real axis. Therefore, the temporal width of the main lobe for the pulse generated by mode-locked laser equals to $4\pi/\Omega$.

After the pulse goes through the first dispersion fiber whose second-order propagation constant equals to $\beta_{2a}$ ($\beta_{2a} < 0$ in our system), length equals to $l_a$ and attenuation coefficient equals to $\alpha_a$. The spectrum of the pulse can be calculated through one-dimensional Nonlinear Schrödinger Equation (NLSE) under sinusoidal steady state regardless of high-order dispersion and nonlinear effects [39], [40].

$$S_2(l_a, \omega - \omega_s) = S_1(0, \omega - \omega_s) \exp\left(j\frac{\beta_{2a}l_a}{2}(\omega - \omega_s)^2\right) e^{-\frac{\alpha_a l_a}{2}}$$

$$= G_\Omega(\omega - \omega_s) \exp\left(j\frac{\beta_{2a}l_a}{2}(\omega - \omega_s)^2\right) e^{-\frac{\alpha_a l_a}{2}} \quad (4)$$

As is seen from (4), the difference between the spectrum of pulse before and after the fiber is only in its phase spectrum. There is no change over the shape of the amplitude spectrum of the pulse when propagating through a dispersion fiber link.

Like the derivation of (3), the time domain signal after the fiber can also be precisely calculated using the Inverse Fourier Transform. In theory, the result cannot be expressed in a closed form. However, since the pulse propagating through the first fiber will undergo a relatively large dispersion, according to the theory of Dispersive Fourier Transform, the time domain pulse can be approximately expressed in a closed form [36].

$$E_2(l_a, t - \beta_{1a}l_a) \cong 2\sqrt{\frac{2\pi}{\beta_{2a}l_a}} \exp\left(-j\frac{(t - \beta_{1a}l_a)^2}{2\beta_{2a}l_a} + j\frac{\pi}{4}\right) S_1\left(0, \frac{(t - \beta_{1a}l_a)}{\beta_{2a}l_a}\right) e^{-\frac{\alpha_a l_a}{2}}$$

$$= 2\sqrt{\frac{2\pi}{\beta_{2a}l_a}} G_{\Omega\beta_{2a}l_a}(t - \beta_{1a}l_a) \exp\left\{j\left(-\frac{(t - \beta_{1a}l_a)^2}{2\beta_{2a}l_a} + \frac{\pi}{4}\right)\right\} e^{-\frac{\alpha_a l_a}{2}} \quad (5)$$

in which $\beta_{1a}$ ensembles the first-order propagation constant when light traveling in the first fiber.

If we establish a new coordinate whose frame moves at the speed of the group velocity of light, then the corresponding transform can be described as

$$T = t - \beta_1 z \qquad (6)$$

in which $\beta_1$ ensembles the first-order propagation constant.

Under this circumstance, (5) can be simplified as

$$E_2(l_a, T) \cong 2\sqrt{\frac{2\pi}{\beta_{2a}l_a}} \exp\left(-j\frac{T^2}{2\beta_{2a}l_a} + j\frac{\pi}{4}\right) S_1\left(0, \frac{T}{\beta_{2a}l_a}\right) e^{-\frac{\alpha_a l_a}{2}}$$

$$= 2\sqrt{\frac{2\pi}{\beta_{2a}l_a}} G_{\Omega\beta_{2a}l_a}(T) \exp\left\{j\left[\frac{-T^2}{2\beta_{2a}l_a} + \frac{\pi}{4}\right]\right\} e^{-\frac{\alpha_a l_a}{2}} \qquad (7)$$

Note that in reality, all signals are real. However, since stationary phase approximation [36] is adapted in Dispersive Fourier Transform, the expression obtained has phase terms. But it is of little significance since all operations in time domain will only be done in its amplitude part.

$$|E_2(l_a, T)| \cong 2\sqrt{\frac{2\pi}{\beta_{2a}l_a}} G_{\Omega\beta_{2a}l_a}(T) e^{-\frac{\alpha_a l_a}{2}} \qquad (8)$$

As is shown in (8), after the first fiber, the original ultrashort $Sa$-shape pulse is broadened to a rectangular-shape pulse with the width of $\Omega\beta_{2a}l_a$. Its shape is the same as the spectrum of the ultrashort pulse after mode-locked laser. In reality, the extent of pulse stretch is limited by the interval time between pulses and the maximum dispersion.

After the pulse is broadened, it will then be modulated with signals. The first modulated signal corresponds to the elements of one row of the weight matrix and the second modulated signal corresponds to the input vector which was turned from the output results from the previous loop. In mathematical analysis, these two modulation processes can be expressed altogether without the loss of generality.

$$E_3(l_a, T) = f(T)E_2(l_a, T) = \sum_{n=1}^{N} \sqrt{w_n x_n} \times$$
$$G_{\Omega\beta_{2a}l_a/N}\left(T - (2n - N - 1)\frac{\Omega\beta_{2a}l_a}{2N}\right) E_2(l_a, T) \qquad (9)$$

in which $w_n$ and $x_n$ represent the $n^{th}$ element in one specific row of the weight matrix and the $n^{th}$ element in the column of the output vector respectively. As can be seen from (9), the original rectangular-shape signal is divided into $N$ slices. The number of slice mainly depends on the number of weight coefficient, the maximum sampling rates of arbitrary waveform generator and the temporal width of the broadened pulse.

Using the property of Fourier Transform, the spectrum of pulse after modulation can be expressed as

$$S_3(l_a, \omega - \omega_s) = \frac{1}{2\pi} F(\omega - \omega_s) \otimes S_2(l_a, \omega - \omega_s)$$

$$= \frac{1}{2\pi} F(\omega - \omega_s) \otimes \left[G_\Omega(\omega - \omega_s) \exp\left(j\frac{\beta_{2a}l_a}{2}(\omega - \omega_s)^2\right)\right] e^{-\frac{\alpha_a l_a}{2}} \qquad (10)$$

in which the operator $\otimes$ represents linear convolution. For continuous form, it refers to linear convolution integral while for discrete form, it refers to linear convolution summation. As is shown in (10), this result cannot be expressed in a closed form since the spectrum function $F(\omega - \omega_s)$ depends on the modulated signal.

By using one dimensional NLSE, the spectrum after the second fiber link whose second-order propagation constant equals to $\beta_{2b}$ (using dispersion compensating fiber whose $\beta_{2b} > 0$ in our system), attenuation coefficient equals to $\alpha_b$ and length equals to $l_b$ can be easily calculated as

$$S_4(l_a + l_b, \omega - \omega_s) = S_3(l_a, \omega - \omega_s) \exp\left(j\frac{\beta_{2b}l_b}{2}(\omega - \omega_s)^2\right) e^{-\frac{\alpha_b l_b}{2}}$$

$$= \frac{1}{2\pi} \exp\left(j\frac{\beta_{2b}l_b}{2}(\omega - \omega_s)^2\right) \times$$
$$\left\{F(\omega - \omega_s) \otimes \left[G_\Omega(\omega - \omega_s) \exp\left(j\frac{\beta_{2a}l_a}{2}(\omega - \omega_s)^2\right)\right]\right\} e^{-\frac{\alpha_a l_a + \alpha_b l_b}{2}} \qquad (11)$$

Its corresponding time-domain signal can be expressed via the Inverse Fourier Transform.

$$E_4(l_a + l_b, T) = \frac{1}{2\pi} \int_{-\infty}^{\infty} S_4(l_a + l_b, \omega - \omega_s) \exp(-j(\omega - \omega_s)T) d\omega$$

$$= \frac{e^{-\frac{\alpha_a l_a + \alpha_b l_b}{2}}}{(2\pi)^2} \int_{-\infty}^{\infty} \exp\left(j\left[\frac{\beta_{2b}l_b}{2}(\omega - \omega_s)^2 - (\omega - \omega_s)T\right]\right) \times$$
$$\left\{F(\omega - \omega_s) \otimes \left[G_\Omega(\omega - \omega_s) \exp\left(j\frac{\beta_{2a}l_a}{2}(\omega - \omega_s)^2\right)\right]\right\} d\omega \qquad (12)$$

Then the compressed optical pulse will be received by a PD.

If the bandwidth of the PD is relatively narrow compared with that of the output signal of ideal PD with infinite bandwidth, the PD will not only serve as an energy transform device, but also a pulse energy accumulator. Under this circumstance, the output signal will reflect the energy of the light pulse it transforms. By using Plancherel Theorem in theory of harmonic analysis [41], this corresponding process can be described and further calculated as

$$\mathscr{E} = \kappa \int_{-\infty}^{\infty} |E_4(l_a + l_b, T)|^2 dT$$

$$= \frac{\kappa}{2\pi} \int_{-\infty}^{\infty} S_4(l_a + l_b, \omega - \omega_s) S_4^*(l_a + l_b, \omega - \omega_s) d\omega$$

$$= \frac{\kappa e^{-\alpha_b l_b}}{2\pi} \int_{-\infty}^{\infty} S_3(l_a, \omega - \omega_s) S_3^*(l_a, \omega - \omega_s) d\omega$$

$$= \kappa e^{-\alpha_b l_b} \int_{-\infty}^{\infty} |E_3(l_a, T)|^2 dT \qquad (13)$$

in which $\mathscr{E}$ represents the accumulated result of the pulse received by PD and $\kappa$ represents the coefficient brought by PD in energy transform and accumulation process. $S^*(\omega)$ represents the complex conjugate of the function $S(\omega)$.

By substituting (8) and (9) into (13) and using the parameters of the function $G$ which is described in (1), we can obtain

$$\mathscr{E} = \kappa e^{-\alpha_b l_b} \int_{-\infty}^{\infty} |f(T)E_2(l_a, T)|^2 dT$$

$$\cong \frac{8\pi\kappa e^{-(\alpha_a l_a + \alpha_b l_b)}}{\beta_{2a}l_a} \int_{-\infty}^{\infty} \sum_{n=1}^{N} w_n x_n G_{\Omega\beta_{2a}l_a/N}^2\left(T - (2n - N - 1)\frac{\Omega\beta_{2a}l_a}{2N}\right) dT$$

$$= \frac{8\pi\kappa e^{-(\alpha_a l_a + \alpha_b l_b)}}{\beta_{2a}l_a} \sum_{n=1}^{N} w_n x_n \int_{(2n-N-1)\Omega\beta_{2a}l_a/2N}^{(2n-N+1)\Omega\beta_{2a}l_a/2N} \frac{4\pi^2}{\Omega^2} dT$$

$$= \frac{32\pi^3 \kappa e^{-(\alpha_a l_a + \alpha_b l_b)}}{N\Omega} \sum_{n=1}^{N} w_n x_n \qquad (14)$$

Then the final result can be obtained as

$$\mathscr{E} = \frac{32\pi^3 \kappa e^{-(\alpha_a l_a + \alpha_b l_b)}}{N\Omega} \sum_{n=1}^{N} w_n x_n = \gamma \sum_{n=1}^{N} w_n x_n \propto \sum_{n=1}^{N} w_n x_n \qquad (15)$$

As is shown in (15), the accumulated result of the pulse reflects the multiplication result of the input vector with one row of the weight matrix. Thus, by using multiple pulses whose number equals to the row of the weight matrix, this structure can implement the multiplication of the input vector and the weight matrix which is shown in the magnified window in Fig. 1.

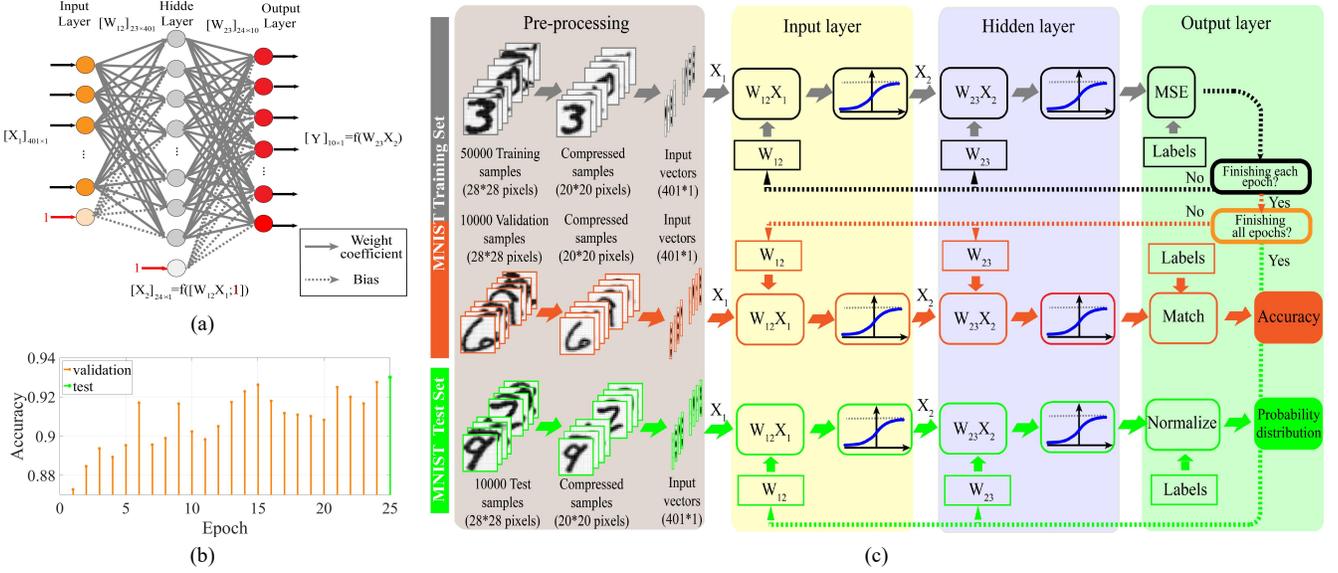

Fig. 2. Structure and training implementation of the theoretical neural network. (a) Theoretical structure. There are 3 layers in total, including one input layer with 400 neurons, one hidden layer with 23 neurons and one output layer with 10 neurons. The last "virtual neuron" at each layer except the output layer is set to be one as to multiply with the bias. (b) Convergence plot of the training algorithm. (c) Training implementation. Pre-processing procedure mainly includes picture compression and input vector generation. In the training session, MSE function is used as loss function to describe the error distance from output to the desired label within each epoch. The 10000 validation samples are used to test the training performance when each epoch is finished. When all epochs are done, 10000 test samples from MNIST test set are used to test the accuracy of classification. For the output layer in the test process, the probability distribution of classification is obtained by using the sigmoid function and then normalizing the output vector.

This expression theoretically proves the statement that by adapting time-stretch method, the core optical structure can implement linear computations in neural networks. Thus, together with nonlinear function simulated by electronic devices, any scale of feedforward neural networks can be implemented by our electro-optical system.

## III. SIMULATION AND RESULTS

Simulation setups and results are discussed in this section. Two procedures must be adapted in order to get the final classification results of TS-NN. Firstly, constructing the theoretical neural network on electronic instruments (the corresponding ENN) and training the network in order to get the optimized weight matrices of the theoretical neural network. Secondly, designing TS-NN and mapping both weight matrices and vectors into modulated signals. The classification task we do here is handwriting digit recognition and all training, validating and testing samples come from MNIST [2].

### A. Constructing and training theoretic neural network

In this procedure, we first construct the theoretical neural network model. For simplicity, the neural network in this article has 3 layers—one input layer, one hidden layer and one output layer. Each original image samples from MNIST digit recognition are re-sized to 20×20 pixels. Therefore, there will be 400 neurons in the input layer. Since we usually add one extra "virtual neuron" which equals to one in each layer except the output layer as to multiply with the corresponding bias later on, in total there are 401 elements in $X_1$. There are 10 neurons in the output layers since all samples will be classified as 0 to 9. In order to determine neurons in the hidden layer and obtain the relatively fast convergence rate, the empirical formula is adapted [42].

$$N_{hidden} = \lfloor \sqrt{N_{input} + N_{output}} \rfloor + N_{emp} \quad (16)$$

where $N_{hidden}$, $N_{input}$ and $N_{output}$ represent the number of neurons in the hidden layer, input layer and output layer respectively. $N_{emp}$ is an empirical integer whose value varies from 0 to 10. Operator $\lfloor \cdot \rfloor$ represents rounding the number to the nearest integer which is no larger than that number. By using the optimization searching algorithm, the best number of neurons in the hidden layer is determined to be 23.

Therefore, as to add one extra "virtual neuron" equaling to one in the hidden layer to multiply with the corresponding bias, there are overall 24 elements in $X_2$. All nonlinear stimulus functions in our neural network system in training processes are non-negative sigmoid function whose value varies from 0 to 1. In order to optimize all weight matrices so as to get the minimum classification error, loss function, training data and training algorithm must be determined respectively.

Loss function is set to be the mean square error (MSE) function since it is easier to calculate when performing error back propagation algorithm. 50000 of all 60000 samples from MNIST training set are used to train our network so as to get the minimum error and the remaining 10000 samples are used as validation samples to validate the performance of classifying after finishing each epoch. Note that the validation samples should be isolated from training samples in order to avoid the data snooping effect. After all training epochs are accomplished, 10000 samples from MNIST test set are used to obtain the final classification accuracy of this theoretical neural network. The Stochastic Gradient Descent (SGD) method is adapted to minimize the total loss and optimize weight matrices [43]. Both structure and training implementation of the theoretical feedforward neural network are described in detail in Fig. 2(a) and Fig. 2(c). The classification accuracy converges in a relatively fast but somewhat fluctuational way to 93% as can be seen from Fig. 2(b). It converges fast since the empirical formula is adapted to determine the number of neurons in the hidden layer in the previous procedure. The fluctuation effect implies the complicated surface of loss function and performances of generalization. Optimized weight matrices are obtained after training procedures.

TABLE I
MAIN PARAMETERS OF TS-NN

| Parameter | Value |
| --- | --- |
| Pulse bandwidth | 2.3nm |
| Pulse period | 20ns |
| Original pulse width | ~1.5ps |
| Peak power of laser output | 50mW |
| Total length of SMF | 170km |
| Dispersion coe. of SMF | 17ps/nm/km |
| Noise figure of EDFA | 4dB |
| AWG sampling rate | 120GSa/s |
| AWG bandwidth | 45GHz |
| AWG vertical resolution | 8 bits |
| Modulator bandwidth | 35GHz |
| Total length of DCF | 28.5km |
| Dispersion coe. of DCF | -100ps/nm/km |
| Peak width of widened pulse | 12ns |
| Responsibility of PD | 0.9 |
| Thermal noise of PD | $1\times10^{-22}$W/Hz |

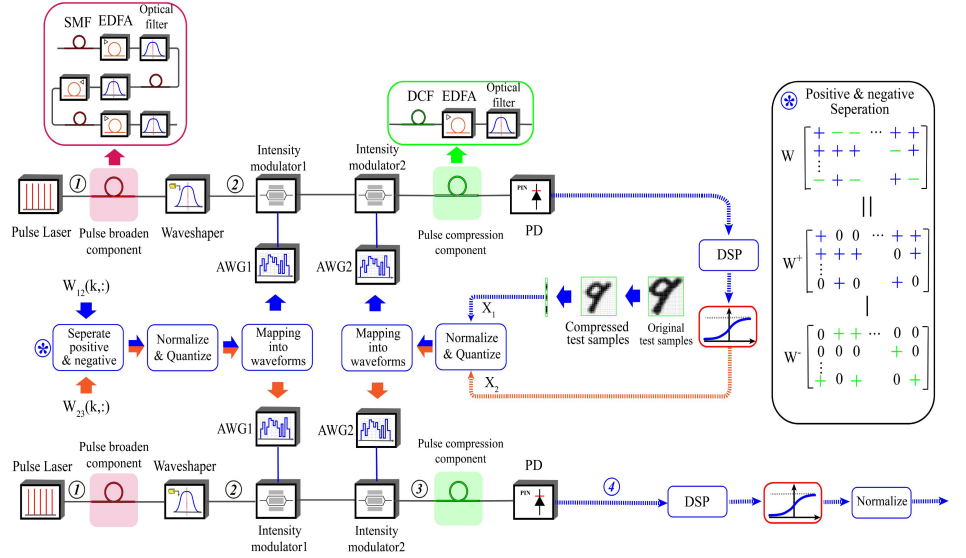

Fig. 3. Structure of simulated TS-NN. The numbers in circles mark important nodes in the system of TS-NN, of which time-domain or frequency-domain signals will be shown in Fig. 4. Black links or arrows represent optical information flow while blue links or arrows ensemble electronic information flow.

*B. Designing TS-NN*

This part of work is to turn the theoretical optimized weight matrices into corresponding modulated signals in our TS-NN system after they were determined in previous steps.

When intensity modulators and single PD are used to transfer information between electronic and photonic devices, only positive coefficients can be transmitted in TS-NN. So the neural network is separated into two parts in each layer so as to calculate positive and negative coefficients respectively. This is because for each layer, negative information only appears in weight matrix which is determined after the training process. Then subtractions can be adapted to get the actual matrix multiplication result. This procedure of separation is described in Fig. 3. Taking this factor into account, in practice, the number of pulses to implement linear computation in each layer should be at least twice as much as the number of rows of the weight matrix.

Compared the simulated implementation of TS-NN which is described by Fig. 3 with the universal structure of TS-NN which is described by Fig. 1, the loop structure is unwrapped as to describe the information flow in a clearer way. As is shown in Fig. 3 and Table I, a mode-locked laser generates periodic pulses light whose bandwidth is 2.3nm and repetition frequency is 50MHz. The peak power of each pulse is set to be 50mW. A single-mode fiber (SMF) (dispersion coefficient is 17ps/nm/km) link is used for pulse stretch. Thus, in order to implement large dispersion, in total up to 170km SMF fiber is adapted in the pulse broaden component. For compensating the power loss during long distance propagation and pulse broaden, three EDFA devices are adapted in this process as is shown in the zoomed-in window of pulse broaden component in Fig. 3. These broadened pulses will then be manipulated by a waveshaper or programmable optical filter to make the peak of pulse flat. Up to 12ns flattened pulse can be obtained after the waveshaper. However, there will still be some fluctuations on the peak of flattened stretched pulse due to imperfections in spectrum manipulation and optical noise which can be seen from in Fig. 4(c).

The modulation component contains two Arbitrary Waveform Generators (AWG) and two Intensity Modulators (IM) whose bandwidth are set to be 35GHz. As shown in Fig. 3 and Fig. 4, intensity modulator 1 is used to modulate waveform which corresponds to each row of $W_{12}^+$, $W_{12}^-$, $W_{23}^+$ or $W_{23}^-$ onto each flattened broadened pulse while intensity modulators 2 is to modulate waveform which corresponds to $X_1$ or $X_2$ onto flattened broadened pulses. Reference pulses as is shown in Fig. 4(e) which are modulated with the multiplication of the maximum absolute value of all weight matrices' elements and elements from input vector are required during the normalization since they can do a great help in aligning the system. Since each row of matrices is separated into positive ones and negative ones, two pulses are needed to conduct one row of original weight matrices as is shown in Fig. 4(e) and Fig. 4(g). All properties of simulated AWGs are based on the Keysight arbitrary waveform generator M8194A whose sampling rate is 120GSa/s, output bandwidth is 45GHz and vertical resolution is 8 bits. Therefore, as shown in Fig. 3, before mapping these vectors and weight matrices into modulated signals, quantization need to be done so as to match the vertical resolution of AWG.

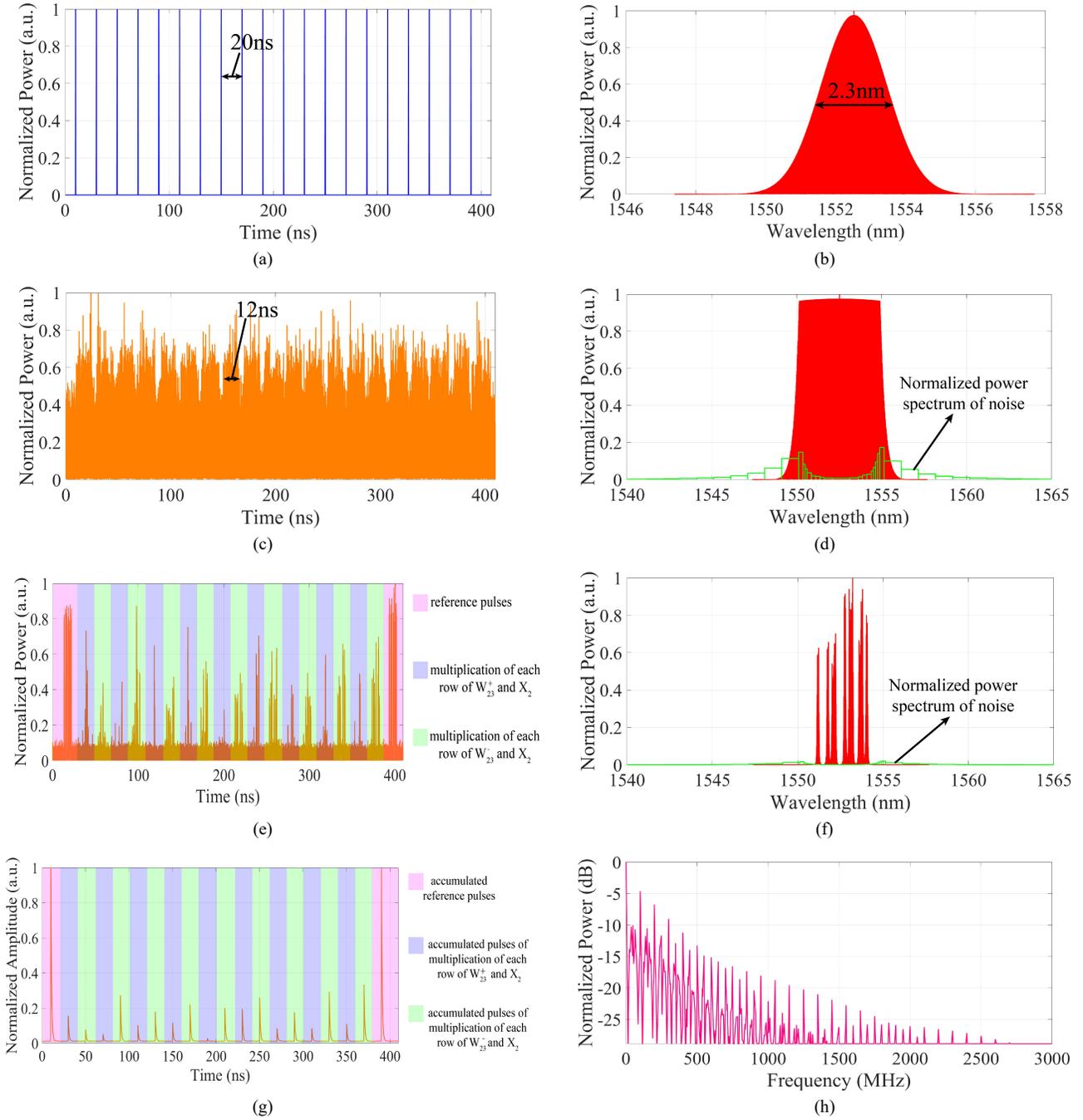

Fig. 4. Waveforms of TS-NN. In total 8 signals from 4 optical nodes in Fig. 3 are plotted. (a) Output pulses of the Laser (①). (b) Output spectrum of the Laser (①). (c) Widened pulses after waveshaper (②). (d) Output spectrum after waveshaper (②). (e) Modulated pulses after intensity modulator 2 (③). (f) Spectrum after intensity modulator 2 (③). (g) Signals after PD (④). (h) Spectrum after PD (④).

For the pulse compression process, the dispersion compensation fiber (DCF) is adapted whose dispersion coefficient equals to -100ps/nm/km and attenuation coefficient is 0.5dB/km. In order to compress the broadened modulated pulse, 28.5km of DCF fiber is adapted in this process. For the PD in the receiver, its bandwidth is set to be 200MHz which is relatively narrow with respect to the bandwidth of RF output of the ideal PD with infinite bandwidth. Therefore, it works both as an energy transform device and accumulation device.

Afterwards, DSP devices like PC will conduct subtractions between the accumulated pulse which is modulated with the multiplication of $W^+$ and X and the accumulated pulse which is modulated with the the multiplication of $W^-$ and X as is shown in the adjacent pulses in blue and green background in Fig. 4(g). Then it will use the signal processing algorithm to extract the actual multiplication of the weight matrix and the input vector from the subtraction results obtained from the first step. All nonlinear functions like non-negative sigmoid functions are implemented by PC. After the multiplication results going through nonlinear functions, they will be rearranged as the input vector which will later be turned into the corresponding modulated signals in the next layer. For nonlinear function in the output layer, the output probability distribution can be obtained by firstly using sigmoid function as the nonlinear function and then normalizing the function outputs.

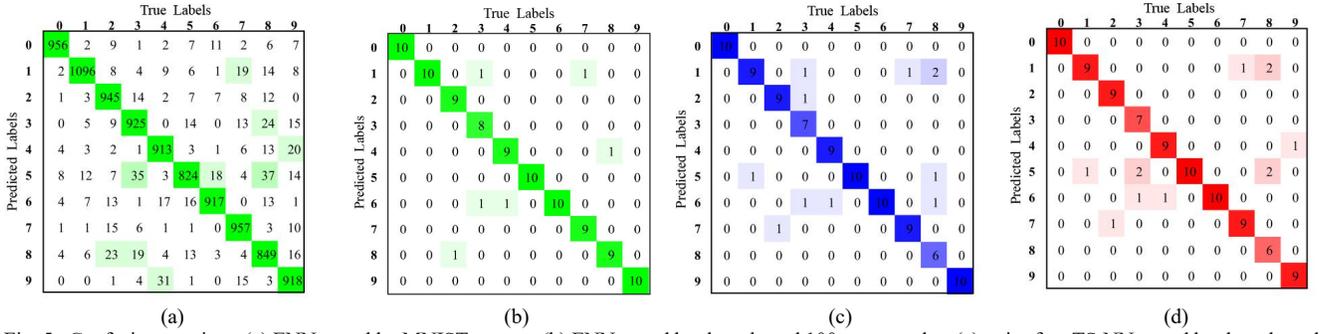

Fig. 5. Confusion matrices. (a) ENN tested by MNIST test set. (b) ENN tested by the selected 100 test samples. (c) noise free TS-NN tested by the selected 100 test samples. (d) TS-NN with noise whose configurations are listed in Table I tested by the selected 100 test samples.

The overall noise in the system is mainly caused by EDFA especially during the pulse broaden process as can be seen from Fig. 4(c). Therefore, in simulation, each optical band-pass filter is adapted after EDFA as to repress the noise. The noise figure of all EDFA in our system is set at 4dB. PD is another noise source and its thermal noise is set at $1\times10^{-22}$W/Hz. However, the noise affects little in our system since the bandwidth of the PD is relatively small enabling it to not only accumulate the energy of pulses, but also repress most of the noise.

*C. Testing and evaluating performances of TS-NN*

In this section, 100 test samples which are randomly picked from MNIST test set (10 test samples for each category from 0 to 9) are used to test the classification performances of three kinds of systems including the corresponding ENN (theoretical neural network), noise free TS-NN and TS-NN with noise. After that, comparisons and analysis will be made. Fig. 5(a) shows the confusion matrix of the corresponding ENN tested by the whole MNIST test set which has been already done in section A. And Fig. 5(b)-(d) show the confusion matrix for each test described above respectively.

As is shown in Fig. 5, the accuracy of the corresponding ENN for total 100 test samples randomly selected from MNIST test set is 94%, which is close to the 93% accuracy for the whole MNIST test set. However, since the fluctuations and other imprecisions occur when mapping weight matrices and vectors on TS-NN, the accuracy falls down to 89%. For TS-NN with noise, the accuracy does not fall down much, which equals to 88%.

More details can also be seen from Fig. 5. Mis-classification usually happens when the pattern distance is relatively small and it is hard for TS-NN to get the exact connection between the test sample and the true digit. This phenomenon can also be seen from the cases where TS-NN mis-classifies 9 into 4 and 7 into 1. For most cases, if mis-classification happens in ENN, then it is of high probable that it will happen in TS-NN as well. In minor cases such as when ENN mis-classfies 8 into 4 while TS-NN classifies it correctly, it may be because the fluctuation effect in flattened stretched pulses can luckily 'compensate' for the wrong weight coefficients obtained from ENN so as to make the final result correct. Other cases show noise, which may not have great impact on classification, or does cause the TS-NN to mis-classify the sample such as the case in which TS-NN with noise mis-classifies 9 into 4 while noise free TS-NN classifies it correctly.

IV. CONCLUSION

In this paper, a novel kind of electro-optical neural networks based on the time-stretch method is proposed. By modulating weight matrices and input vectors onto flattened stretched pulses, a 3-layer electro-optical feedforward neural networks with 400, 23, 10 neurons in the input, hidden and output layer respectively can be implemented. This electro-optical neural network reaches up to 88% accuracy under relatively strong noise brought by both optical amplifiers and PD devices in classifying handwriting digits.

Several advantages can be seen once this structure is adapted. Since our electro-optical system implements feedforward neural networks in a serial way via time stretch method, it shows almost no extra cost when finishing more complicated and large scale tasks. It is relatively easier for researchers to align and adjust the system since the number of parameters of the system will not expand severely as the scale of neural networks increases. The most important advantage of this serial way to transmit and calculate information is that the traditional electronic neural networks test sets like MNIST handwriting digit recognition set which are impossible to be operated on most traditional ONN structures before can be operated in our electro-optical neural network system. Besides, due to the use of electronic devices, nonlinear function can be implemented precisely. Also, the ability of reconfiguration of our system is better compared with conventional structures of optical feedforward neural networks such as $D^2$NN or chip-integrated neural network since the only effort to reconfigure the neural networks is to change the modulated weight coefficients and number of loops. Thanks to the narrow bandwidth of PD, the noise has little impact on TS-NN which can be clearly seen by comparing Fig. 5(d) with Fig. 5(c). Furthermore, the system has a clear correspondence between the theoretical neural network model and the actual optical and electronic component parameters. Each signal or device parameter in our system can find clear relations with the corresponding signal and parameter of the theoretical feedforward neural network model. Although our scheme is numerically simulated with discrete optical and electronic devices, this method can be feasible to combine with many optoelectronic integration technologies to reduce the complexity and difficulty of the system and improve usability.